%% file: paper.tex
\documentclass[runningheads]{llncs}
\usepackage[T1]{fontenc}
\usepackage{graphicx}

\usepackage{amssymb}

\usepackage[utf8]{inputenc}
\usepackage{listings}

\usepackage{todonotes}
\begin{document}
\title{Design Principles of Dynamic Resource Management for High-Performance Parallel Programming Models}

\titlerunning{Design Principles of DRM for HPC Parallel Programming Models}
%
%
\author{Dominik Huber\inst{1}\orcidID{0000-0001-9696-9382} \and
Martin Schreiber\inst{1,2}\orcidID{0000-0002-2390-6716} \and
Martin Schulz\inst{1}\orcidID{0000-0001-9013-435X} \and
Howard Pritchard\inst{3}\orcidID{0000-0003-1969-0403} \and
Daniel Holmes\inst{4}\orcidID{0000-0002-2776-2609}}
\authorrunning{D. Huber et al.}
%
\institute{Technical University of Munich, Boltzmannstr. 3, 85748 Garching, Germany
\email{\{domi.huber,martin.w.j.schulz\}@tum.de} \and
Université Grenoble Alpes, 700 Avenue Centrale, 38400 Grenoble, France
\email{martin.schreiber@univ-grenoble-alpes.fr} \and
Los Alamos National Laboratory, Los Alamos, NM 87545, USA \\
\email{howardp@lanl.gov} \and
Intel Corporation, 2200 Mission College Blvd. Santa Clara, CA 95054-1549, USA
\email{daniel.john.holmes@intel.com}
}

\maketitle              

\input{chapters/00_abstract.tex}

\input{chapters/01_introduction.tex}
\input{chapters/02_paradigms.tex}

\input{chapters/03_realization.tex}

\input{chapters/04_discussion.tex}
\input{chapters/05_related_work.tex}

\input{chapters/06_summary.tex}

\input{chapters/99_appendix_history.tex}

\begin{credits}
\subsubsection{\ackname} This project has received funding
from the Federal Ministry of Education and Research and the
European HPC Joint Undertaking (JU) under grant agree-
ment No 955701, Time-X, No 955606, DEEP-SEA and No grant agreement No 956560, REGALE. The
JU receives support from the European Unions Horizon 2020
research and innovation programme and Belgium, France,
Germany, Switzerland.

\subsubsection{\discintname}
The authors have no competing interests to declare that are
relevant to the content of this article.
\end{credits}
%
%

\bibliographystyle{splncs04}
\bibliography{paper.bib}

\end{document}

%% file: chapters/00_abstract.tex
\begin{abstract}
With Dynamic Resource Management (DRM) the resources assigned to a job can be changed dynamically during its execution.
From the system's perspective, DRM opens a new level of flexibility in resource allocation and job scheduling and therefore has the potential to improve system efficiency metrics such as the utilization rate, job throughput, energy efficiency, and responsiveness, and is key for efficiency in scenarios like urgent computing jobs. 
From the application perspective, users can tailor the resources they request to their needs offering potential optimizations in queuing time or charged costs. 


Despite these obvious advantages and many attempts over the last decade to establish DRM in HPC, it remains a concept discussed in academia rather than being successfully deployed on production systems.
This stems from the fact that support for DRM 
requires changes in all the layers of the HPC system software stack including applications, programming models, process managers, and resource management software, as well as an extensive and holistic co-design process to establish new techniques and policies for scheduling and resource optimization, which can get arbitrarily complex.


In this work, we therefore start with the assumption that resources are
accessible by processes executed either on them (e.g., on CPU) or controlling
them (e.g., GPU-offloading). Then, the overall DRM problem can be decomposed
into dynamic process management (DPM) and dynamic resource mapping or
allocation (DRA). The former determines which processes (or which change in processes) must be managed and the latter identifies the resources where they will be executed.
The interfaces for such DPM/DPA in these layers need to be standardized, which requires a careful design to be interoperable while providing high flexibility.
In this paper, we will survey existing approaches and highlight that most attempts so far fall short of these requirements.

Based on these observations, we propose design principles for DRM and argue why these principles can form the basis of a holistic approach to dynamic resource management in HPC and AI.
We also discuss a possible realization of these design principles based on an actual prototype using the Message Passing Interface (MPI).

\keywords{Dynamic Resource Management  \and Parallel Programming Models \and High Performance Computing}
\end{abstract}

%% file: chapters/01_introduction.tex
\section{Introduction}
\label{sec:introduction}
Traditionally, static resource management is used on HPC systems, i.e., the resources assigned to a workload are fixed during their execution.
Dynamic Resource Management (DRM) allows for dynamically changing the resources assigned to workloads at run time.
DRM has gained considerable interest over the last years as it has the potential to provide many benefits to providers of HPC systems as well as their users.

\subsection{Benefits of Dynamic Resource Management}
From the system provider perspective, DRM increases the scheduling flexibility available to the system, which then can lead to an improvement of several system metrics such as increasing energy efficiency, job throughput and utilization rate~\cite{prabhakaran16}.
In addition, it allows the resource manager to better address dynamic changes in the system, such as network contention, proactive fault tolerance and dynamic power management, and to flexibly react to dynamic workload shifts, like the arrival of urgent computing jobs. 

From the user perspective, the increased system throughput can lead to reductions of the average waiting and turnaround time of a job~\cite{ozden23}.
This translates to a reduction in "time to insight", the duration it takes researchers and scientists to obtain meaningful results or insights from their computational simulations or analyses. Moreover, it can increase the efficiency of some workloads by addressing their inherent dynamicity, such as Adaptive Mesh Refinement methods~\cite{schreiber_invasive_2015}.

An indirect benefit of improved efficiency is also increased 
energy efficiency, which  
addresses two of the four pillars described in ``The 4 Pillar Framework for energy efficient HPC data centers''~\cite{wilde13}. With the ever-growing demand for HPC, improving the energy efficiency of HPC data centers is essential.

Beyond this, a more elastic resource model for HPC can also facilitate the shift towards Converged Computing \cite{milroy21}, i.e., the use of HPC applications on more dynamically scheduled cloud resources and the deployment of cloud-like applications on HPC resources.

\subsection{Terminology}
DRM contributes to the main goal of \textit{HPC resource management software}: the 
assignment of \textit{HPC resources} to \textit{HPC execution concepts} to achieve efficient utilization of the resources provided by the system. 
As HPC systems continue to evolve in complexity and accommodate increasingly 
diverse use cases with diverse resource requirement scenarios, the conventional terminology employed to describe the concepts needed for DRM has grown more ambiguous. We, therefore, first
introduce the terminology we use in this work to describe the different concepts on HPC systems, as well as a model of their relations (Fig.~\ref{fig:layers}). 


\subsubsection{HPC Execution Concepts}
HPC execution concepts are concepts that describe the execution of computational work using HPC resources.
There are several \textit{execution concepts} of different granularity:

\textbf{Process:} A process is a 
fine-grained execution concept. Processes execute instances of a single executable with its own address space and execution context and possibly multiple execution threads. Processes are executed 
on resources (e.g. a CPU core) and have access to certain \textit{HPC resources} (e.g. to multiple CPU cores using threads or to GPUs using offloading). 
Processes are a way to exploit parallelism on HPC resources. 

\textbf{Application:} In this work we assume an application instance 
to be a single executable executed 
in parallel 
by one or more processes. Applications have access to the union of the resources associated 
with the processes executing the application. Applications and 
application libraries usually 
are expressed using 
parallel programming models, such as MPI, to exploit and manage 
the parallelism present when used with multiple processes.

\textbf{Job}: A job consists of one or more application instances executed as a single invocation. It has access to the union of the resources accessed by the application instances it consists of. The applications composed in a single job often share a common communication context, e.g. a global MPI communicator.

\textbf{Workflow:} A workflow typically consists of multiple interdependent jobs running 
concurrently
and/or sequentially.


\subsubsection{HPC Resources}
\textit{HPC resources} are the essential components and capabilities that contribute to the 
execution of execution concepts on HPC systems. Multiple types of resources exist, including:

\textbf{Compute Resources}: Compute resources are used by 
execution concepts to carry out computations. 
This includes CPU cores, GPU execution engines, cores on SmartNICs, as well as processing elements on other accelerators.

\textbf{Storage Resources}: Storage resources are used by execution concepts to store and load data required for the computations. This includes, for instance, memory (typically RAM), persistent storage as well as more structured storage like databases.

\textbf{Indirect Resources}: Indirect resources are resources that are 
not directly used by the execution concepts, but are
associated with compute and storage resources. 
For instance, network resources, including node-local and internode networks,
 connect different compute and/or storage resources.
Energy resources, such as energy budget, 
and cooling systems are associated with the operation of compute and storage resources. 

Reactivity and predictability are qualities associated with computing resources, storage resources and system services.


\subsubsection{HPC Resource Management Software}
The task of HPC resource management software is the assignment of HPC resources to execution concepts to facilitate their execution.
According to the different types of HPC resources and execution concepts, the term HPC resource management software can refer to a wide range of different software components embedded in the HPC software stack. This includes:

\textbf{System-level Resource Manager:} The system-level resource manager manages the submission and scheduling of 
workflows and assigns resources to workflows through resource allocations. 
It is usually configured directly by system administrators to align with specific optimization objectives and policies desired by the 
hosting entity.

\textbf{Process Manager/Parallel Programming Model Runtime Environment:} A process manager is usually running inside of an 
allocation 
provided by the system-level resource manager and is responsible for the job execution. It manages the lifecycle of processes, including launch, termination and failure detection. The process manager also often provides a runtime environment for the employed parallel programming model, such as MPI.

\textbf{Application-specific resource management libraries:} Many HPC applications use libraries for the management of application resources and load-balancing, which are specialized for the particular type of application. These libraries often serve as a shim between the application and the runtime environment to ease the complexity of resource management for application developers. 

\subsection{The Need for Dynamic Resource Management Interfaces}
\begin{figure}
	\centering
	\includegraphics[width=\textwidth]{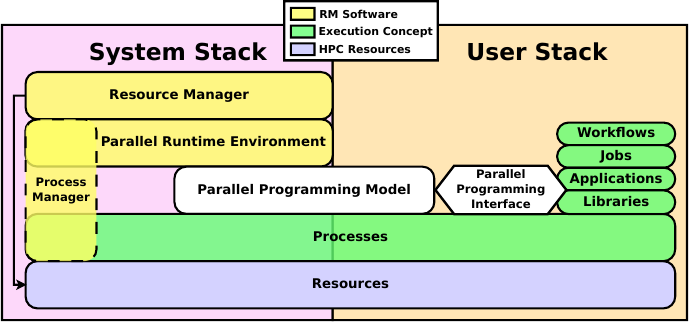}
	\caption{\textbf{Model for the relations between execution concepts, resources and resource management software on HPC systems.}
The access to \textit{HPC resources} (bottom) is controlled by the \textit{resource manager} (top).
\textit{Applications} (right) require access to HPC resources to execute their workload.
Applications access resources through application \textit{processes}, which are typically abstracted by a 
\textit{parallel programming model} and exposed to the application (directly or through application libraries) via a 
\textit{parallel programming interface}.
Processes are managed by the \textit{process manager} (left) and exposed to the parallel programming model by the \textit{parallel runtime environment}.
Dynamic resource management requires coordination between applications and the resource manager throughout multiple software layers: The programming model (interface), the parallel runtime environment and the process manager.}
	\label{fig:layers}
\end{figure}

When shifting from static resource management to dynamic resource management, the concepts mentioned in the previous section (resources, execution concepts and resource management software) are all subject to dynamicity, i.e. dynamically changing associations with resources.
As a consequence, introducing dynamicity requires sophisticated coordination between the different software components, and this, in turn, requires significant adaptions of many components of HPC system software stack (Fig.~\ref{fig:layers}).
This calls for a flexible, generic, and consistent design for dynamic resource management to holistically optimize resource usage on HPC systems and to enable interoperability between the diverse software components.
First attempts at defining interfaces for dynamic resource management achieved some degree of dynamicity for certain scenarios as outlined in more detail in Sec.~\ref{sec:related_work_hpc}. 
However, these interfaces do not fulfill the requirements set forth above, thus impeding the common adoption of dynamic resource management in production HPC systems. 

\subsection{Approach and Structure of this Work}
Our work is based on the assumption that the overall DRM problem can
be decomposed into dynamic resource mapping/allocation (DRA) from the resource management software perspective and dynamic process management (DPM) from the application perspective. 
In this situation, dynamic process changes such as replacing existing processes, adding processes for post-processing, increasing the parallelism of a data parallel task, and rearranging process layouts have a direct relation to dynamic changes of resource allocations.
Therefore, it is essential to design an interface for parallel programming models in HPC based on a co-design approach that considers the application programming perspective and the system resource management perspective.
In particular, the design should contribute to both, the flexibility for programmers to express and manage the specific dynamic behavior of their applications and the flexibility of resource management software to optimize the utilization of resources of the HPC system. 

The central contribution of this work is the introduction of some fundamental 
design principles that we consider to be essential for the design of such a flexible, generic, and consistent interface for dynamic resource management in HPC.

In the following section, we give an extensive description of these design principles.
Subsequently, in Sec.~\ref{sec:proposal} we discuss the application of these design principles to the widely used standard interfaces MPI and PMIx, and provide a concrete realization of these principles.
In Sec.~\ref{sec:discussion} we discuss our experiences with a prototype implementation of these interfaces and point out future research directions.
An overview of related work of dynamic resources in Cloud Computing and HPC is given in Sec.~\ref{sec:related_work}, where we highlight in particular the need for a flexible standardized interface for dynamic resources in HPC.
Finally, we give a summary in Sec.~\ref{sec:summary} and provide our sources of inspiration in Appendix~\ref{sec:appendix}.   ´

%% file: chapters/02_paradigms.tex
\section{Principle-driven Design of a Dynamic Resource Allocation Interface}
\label{sec:principles}

\begin{figure}[hb]
\label{fig:principles}
\includegraphics[width=\linewidth]{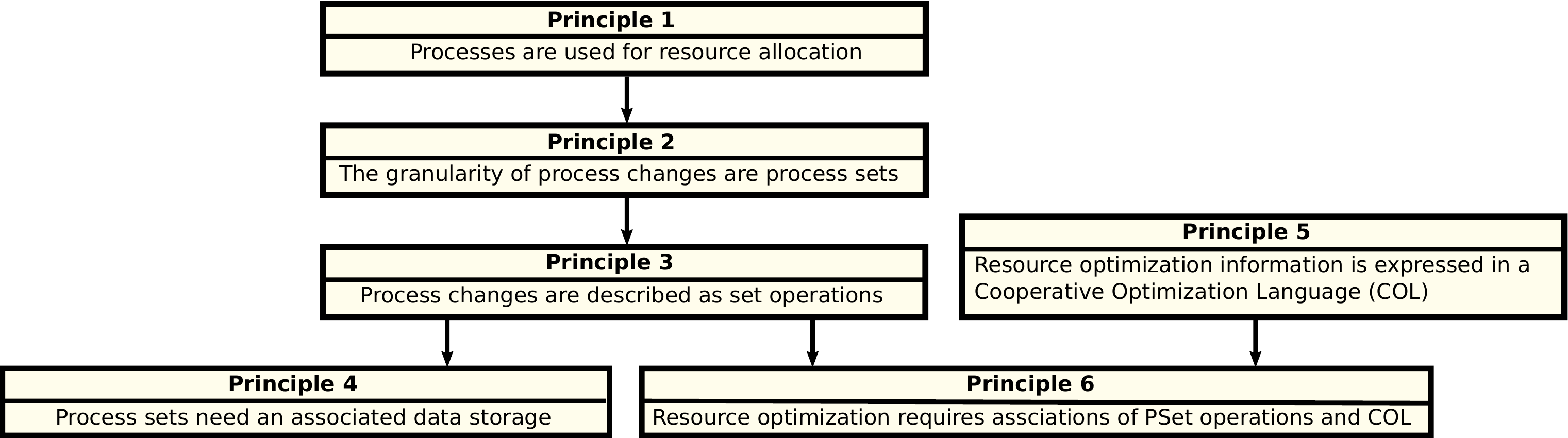}
\caption{Design principles for a dynamic resource allocation interface.}
\label{fig:principles}
\end{figure}

This section is motivated by the desire to establish a standard for a dynamic resource allocation interface embedded in parallel programming models for HPC.
There are many ways to develop such a new interface. To avoid getting lost in technical details we motivate our work with principles.
These principles should be interpreted as the bare minimum, which has to be supported by an interface.
If not, the created interfaces fail to be generic enough to establish a new standard.

As part of our goal to create a new very generic approach, we try to avoid the term ``resource request'' since a request is implicitly included by the optimization problem where, e.g., a particular optimization constraint of a request can be included.

Fig.~\ref{fig:principles} gives an overview of the principles and their relations, which we describe in more detail in the following subsections.

\subsection{Principle 1: Processes are used for resource allocation}
\textbf{Brief description:}
As a separation of concerns between dynamic resource allocation (DRA) and dynamic process management (DPM) the abstraction level for an application interface is given by processes that use resources indirectly.
Thus, resources are never exposed to applications directly, but indirectly via processes.
Processes are then associated with resources by the resource management software based on optimization information, which can e.g. be enforced by CPU masks.

\textbf{Rationale:}
HPC Resources are used by application processes to execute computations.
Therefore, to access resources, they first have to be affiliated with processes that can then use these resources to, e.g., execute a program's thread,
perform communication, or offload kernels to GPUs.

Hence, the resource management software accomplishes dynamic resource allocation (DRA) by creating and terminating processes and making them behave in a particular (e.g., restrictive) way towards resources. Processes could, for instance, be restricted to particular CPU cores (affinity masks) or particular GPU resources. Optimal association between processes and HPC resources according to a global optimization objective is often system-specific and therefore the responsibility of the resource management software. 

As a separation of concerns, the application on the other hand should not refer to resources directly, but instead, processes should be used as an abstraction level of an application interface for dynamic resources. Therefore, this interface is a dynamic process management (DPM) interface targeting dynamic process changes (creation, termination, and reference of processes). 

\subsection{Principle 2: The granularity of process changes are process sets.}

\textbf{Definition ``process set'' (PSet):} A PSet is an ordered set of processes
 with a unique identifier. Thus, the concept is identical to \textit{MPI Process sets} except for requiring the (implicit) ordering.

\textbf{Definition ``$0$-PSet'':} A special empty PSet which always exists.

\textbf{Brief description:}
A PSet allows for a flexible granularity of references to processes for dynamic process management purposes. 

\textbf{Rationale:}
A generic approach for DRM should not restrict the granularity of process changes to, e.g., changing the size of one HPC job.
A PSet granularity can be finer, equal, or coarser than that of an HPC job.
Thus, PSets provide a flexible granularity of reference for dynamic process management.

A PSet needs a unique identifier since the information of PSets needs to be known throughout different layers of the software stack.
This enables referencing the processes (and associated resources) through different layers without explicitly listing them, which would be inefficient in large-scale scenarios. 

\subsection{Principle 3: Process changes are described as set operations between PSets}

\textbf{Definition ``PSet operation'':} A PSet operation is an operation on PSets as input that produces a set of output PSets based on a well-defined rule.
A PSet operation can be concretized with parameters, restrictions, and requirements.

\textbf{Brief description:}
The creation, change, or removal of PSets has to be expressed as set operations with respect to existing PSets or the $0$-PSet.
This also ensures a relationship between resource changes and PSets, which needs to be preserved as discussed below.

\textbf{Rationale:}
A change of processes can have different requirements depending on the particular context.
For instance, a job can encompass different algorithms, such as in ocean-atmosphere coupled simulations.
In such cases, different subsets of the processes running in a job or application can have different requirements and implications on resource changes (e.g., data locality).
Therefore, it must be possible to \textit{relate} process changes to PSets.
Describing process changes as set operations allows them to be related to particular PSets, which are the input/output of the operation.

Changing processes can fulfill different purposes such as replacing existing ones, adding processes for post-processing, increasing the parallelism of a data parallel task, etc.
Set operations provide an abstraction that allows expressing such use cases in a generic, well-defined way.

\subsection{Principle 4: PSets need an associated data storage}
\textbf{Brief description:}
A global data storage to asynchronously store and retrieve data associated with a given PSet.

\textbf{Rationale:}
Processes require application-specific information about PSets and PSet operations in a dynamic environment, e.g., to relate PSets with particular application phases, or application code paths that cannot be known apriori. 
Synchronous data exchange would require direct communication channels between the processes and would enforce strict limitations on the order and timing of message exchange. Thus, to be applicable in a highly dynamic environment, a more generic approach for providing and accessing relevant PSet-related information is required. A global data storage allows processes to provide and access PSet-specific information asynchronously, without direct communication channels between processes.

\subsection{Principle 5: Resource optimization information is expressed in a Cooperative Optimization Language (COL)}

\textbf{Definition ``Cooperative Optimization Language (COL)''}
A Cooperative Optimization Language expresses optimization information for resource management in a \textit{local}, \textit{cooperative}, and \textit{language-based} way.

\textbf{Brief description:}
\textit{Cooperative} refers to the non-competitiveness of resource allocation this approach aims, i.e. different applications contribute to a global optimization goal.

\textit{Local} refers to the local perspective, e.g., a single PSet, used when expressing optimization information.
In particular, no global assumptions, such as guessing the system state, should be 
made, which would lead to a competitive approach.
This locality also means that the provided optimization information does not provide by itself a well-defined optimization problem due to the lack of, e.g., the objective function.

\textit{Language-based} refers to a new (domain-specific) language and its representation required to express this complex optimization information.

\textbf{Rationale:}
DRM requires optimization information from different sources to optimize the assignment of HPC resources to HPC computation concepts according to a global optimization objective.
Expressing resource optimization information as objects containing a description in the COL allows gathering different COL objects and combining them as a global optimization problem.
As a result, a globally optimal solution 
for the resource assignments of the system can be approximated, respecting the local optimization information.

%
%
%
Since the COL plays a key role in optimizing dynamic resource management, we provide further discussion about it as follows:

\subsubsection{``Local'' aspect of the COL}


The ``local'' aspect of the COL means that optimization information should be provided by the application (or PSet) only for itself and should not contain any assumptions on the global state of the system.
I.e. it only contains implicit requests for resource changes through a description of its local performance behavior instead of explicit requests (unless necessary e.g. for urgent computing scenarios). 

Such information can relate, e.g., to the throughput (time to solution) of the PSet or application,
 to the energy consumption, to the memory requirements or the network requirements. It is provided either directly by an application, indirectly by monitoring 
 data, or is based on pre-processed data such as past execution traces.
We refer to this as local information since it's only specific to an application or its PSet, hence a local perspective.

Based on this local information, the dynamic scheduler can make decisions for a global optimization target. This decision is based on local information and the optimization policies given by the system (operators of the compute cluster, e.g., optimize for throughput or power consumption).


\subsubsection{``Cooperative'' aspect of COL}

As explained before, due to the local aspect of the COL, applications express information about their local resource requirements.
At the same time, system-generated data (monitoring, outside decisions, etc.) has to be taken into account.

This leads to the requirement of a cooperative optimization approach. In this context, cooperative means that different actors contribute to a global goal by providing local optimization information and possibly local objectives.
By providing local optimization information such as performance models and resource requirements, different actors contribute to the global optimization of resource assignments in the system, which in turn can lead to better 
local resource assignments.

Such a cooperative approach is in particular important to avoid actively bargaining and competing for resources,
which can easily lead to instabilities due to nonlinear effects.


\subsubsection{``Language'' aspect of COL}

Altogether, the complexity of the optimization information gets extremely large, as it needs to convey e.g. information about application behaviors in different dynamic scenarios and specialized hardware might be required to evaluate it.

Hence, the cooperative resource requests need to be done in a way that allows their evaluation on different computer architectures.
Thus, the language aspect of COL requires the information to be given in a certain language.
This language should be based, e.g., on an intermediate representation (IR) since it can be required to evaluate the COL objects on other hardware than where the PSet is running.
Moreover, the optimizer itself might not require all or only a fraction of the COL where an IR allows such an important step.

\subsection{Principle 6: Resource optimizations require an association between PSet operations and COL objects}

\textbf{Brief description:}
Providing new information for resource optimization requires the pair of a PSet operation and an associated COL object.

\textbf{Rationale:}
A PSet operation on itself only describes \textit{a type of process change} that could be applied to particular application PSets. From an application (programmer) perspective, a PSet operation allows the definition and handling of particular changes in the set of PSets that are available for workload execution.
As a separation of concern, a COL object describes \textit{parameters}, such as particular constraints or requirements, and their \textit{impact} on the workload execution of process sets.
Thus, to achieve both, programmability and resource optimization, a PSet operation needs to be associated with a COL object.

%% file: chapters/03_realization.tex
\section{Concrete Realization of the Design Principles}
\label{sec:proposal}
In this Section, we illustrate how the design principles from the previous section could be realized.
To this end, we formulate extensions to two interfaces commonly used in HPC: MPI and PMIx.

\subsection{MPI}
MPI is the de facto standard for inter-process communication in HPC.
Therefore, MPI should provide an interface for dynamic resources to applications.
In the following subsections, we provide insights into the realization of our design principles in MPI and propose an exemplary MPI interface.
\subsubsection{Possible realization of design principle in MPI}
~

\textbf{Principle 1:}
MPI is based on MPI Processes. Thus, following principle 1 \textit{MPI processes are used for resource allocation}.
MPI processes are associated with resources such as CPU cores through mapping/pinning or other means.

\textbf{Principle 2:}
The MPI Standard 4.0 introduced MPI Process Sets (PSets) as immutable, unordered sets of MPI processes with a unique identifier in the URI format.
The list of MPI processes can be queried with \lstinline{MPI_Session_get_num_psets} and \lstinline{MPI_Sesison_get_nth_pset}.
MPI PSets can be used to derive MPI Groups with \lstinline{MPI_Group_from_session_pset}.

Following principle 2, \textit{the granularity of MPI process changes should be MPI process sets}.

\textbf{Principle 3:}
To follow principle 3 \textit{MPI process changes would be described as set operations and relations between MPI PSets}.
This requires new MPI routines to express set operations.
As a consequence of MPI PSets being immutable, the output sets of a set operation on MPI PSets have to be MPI PSets with a new, unique URI or the 0-PSet.

\textbf{Principle 4:}
Since MPI is about communication rather than resource optimization, as a separation of concerns the contents of the COL should not be part of the MPI specification.
However, MPI would need to be aware of the COL object as an integral part of process changes and therefore provide an appropriate interface to pass on COL objects (e.g. as a byte buffer).

\textbf{Principle 5:}
As Principle 5 mandates the association between PSet operations and COL objects, MPI needs to provide functionality for associating MPI PSet operations with COL objects. Thus, MPI needs to be able to handle COL objects passed by the user and associate them with a given MPI PSet operation.

\textbf{Principle 6:}
As mandated by \textit{Principle 6} process sets need an associated data storage to allow the association of user data with process sets. Thus, MPI needs to provide functionalities to allow MPI processes to publish and look up global data and associate this data with the URI of a given MPI Process set.

\subsubsection{Interface Proposal}
\begin{figure}[t]
\centering
\includegraphics[width=0.9\textwidth]{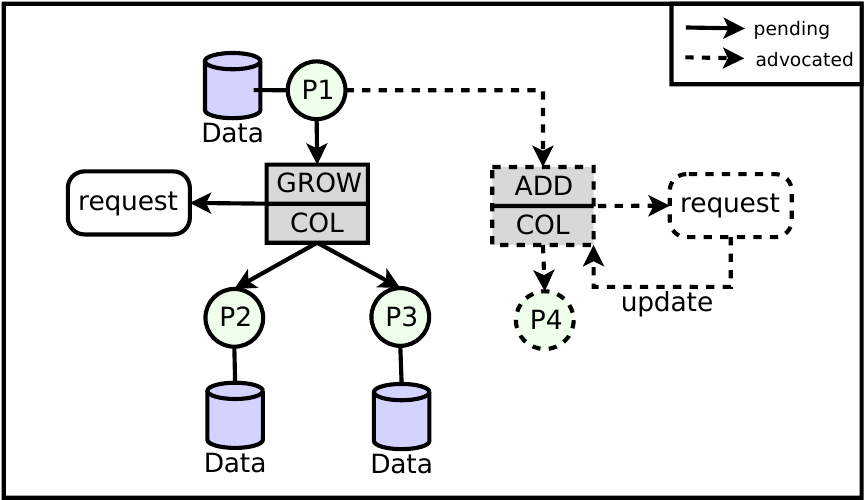}
\caption{\textbf{Illustration of set operations in MPI.} The figure shows an example of two set operations, GROW and ADD advocated by the application via calls to \lstinline{MPI_Session_dyn_psetop}. In both calls the same input PSet was specified, i.e. PSet P1, and for each operation a COL object has been passed to provide optimization information. The call returns an MPI Request, which can be used in a call to \lstinline{MPI_Test} or \lstinline{MPI_Wait} to check if the operation has been executed, or passed to a subsequent \lstinline{MPI_Session_dyn_psetop} call to update the optimization information. In this example the GROW operation has already been executed by the resource manager, i.e. the output PSets have been created and the operation is now \textit{pending}. Pending operations can be queried by all processes using \lstinline{MPI_Session_dyn_query_psetop} until an explicit call to \lstinline{MPI_Session_dyn_finalize_psetop}.
Each PSet has an associated data store where data can be stored and retrieved via the \lstinline{MPI_Session_publish/lookup_data}.
This example could for instance represent a job where the processes in PSet P1 run a simulation. Processes could be added dynamically to the simulation (GROW) or started separately to run some in-situ task (ADD).}
\label{fig:setops}      
\end{figure}

Based on the observations from the previous subsection, we propose the following possible MPI interface which follows the design principles. An illustration is given in Fig.~\ref{fig:setops}. Note that this interface represents only one of multiple possible realizations of the design principles in MPI.

\textbf{Supported MPI PSet operations}
We propose the following MPI constants and PSet operations to be supported by MPI:
\begin{itemize}
\item \lstinline{MPI_PSETOP_NULL}: noop. Used as output to indicate that no PSet operation is pending.
\item \lstinline{MPI_PSETOP_ADD}: Takes $n>0$ input Psets and creates $m$ disjoint output PSets containing new processes.
\item \lstinline{MPI_PSETOP_SUB:} Takes $n>0$ input Psets and creates $m>0$ output PSet containing processes supposed to terminate. The output Psets build a subset of the set of input PSets.
\item \lstinline{MPI_PSETOP_SPLIT:} Takes one input PSet and creates $m>0$ output PSets which are disjoint subsets of the input PSet.
\item \lstinline{MPI_PSETOP_UNION:} Takes $n$ input PSets and creates one output PSet which contains the \textit{union} of the processes in the provided input PSets.
\item \lstinline{MPI_PSETOP_DIFFERENCE:} Takes $n$ input PSets and creates one output PSet which contains the \textit{difference} of the processes in the provided input PSets.
\item \lstinline{MPI_PSETOP_INTERSECTION:} Takes $n$ input PSets and creates one output PSet which contains the \textit{intersection} of the processes in the provided input PSets.
\item \lstinline{MPI_PSETOP_GROW:} Takes one input PSet $S_{in}$ and creates two output PSets: $S_{new}$ containg new processes and $S_{grow} = S_{in} \cup S_{delta}$.
\item \lstinline{MPI_PSETOP_SHRINK:} Takes one input PSet $S_{in}$ and creates two output PSets: $S_{del}$ containg processes supposed to terminate and $S_{shrink} = S_{in} \cap S_{delta}$.
\item \lstinline{MPI_PSETOP_REPLACE:} Takes one input Pset $S_{in}$ and creates 3 output PSets: $S_{new}$ containing new processes, $S_{del}$ contaning processes to be terminated and $S_{repl}: S_{in} \cup S_{new} \cap S_{del}$.
\end{itemize}

\textbf{Specifying MPI PSet operations}:
The procedure \lstinline{MPI_(I)Session_dyn_psetop} is used by MPI processes to specify a pair of an MPI PSet operation and a COL object and transmits this information to the MPI runtime environment.
Upon completion, the names of the resulting output PSets are returned and the PSet operation is considered as \textit{pending}.
Listing \ref{lst:mpi_psetop} provides the signature and description for a blocking and a non-blocking version of this procedure.

\begin{lstlisting}[frame=single, basicstyle=\scriptsize, language=C, label={lst:mpi_psetop}, escapeinside={<@}{@>}, caption={Procedure for transmitting MPI process set operations to the Resource Manager}]
int MPI_(I)Session_dyn_psetop(
  MPI_Session session    // IN: session handle
  int op_in              // IN: The requested set operation 
  char **input_psets     // IN: Names of the input Psets
  int ninput_psets       // IN: Number of input Psets
  void *col_buf        	 // IN: buffer containg the COL object
  int col_buf_len	 // IN: size of col_buf
  int *op_out		 // OUT: The operation that was executed
  char ***output_psets   // OUT: Names of the ouput Psets
  int *n_output_psets    // OUT: Number of output Psets
  (MPI_Request *request) // INOUT: An MPI Request (non-blocking version)
)

<@\textbf{Description:}@> Transmits the specified pair of PSet Operation and COL object 
to the MPI runtime environment. In the <@case@> the provided request 
handle != MPI_REQUEST_NULL, the provided COL object will be associated 
the PSet operation that returned the request handle. If successful, the 
function allocates an array of n_output PSet names. It is the callers 
responsibility to free the allocated memory.
\end{lstlisting}

\textbf{Querying pending MPI PSet operations}:
The procedure \lstinline{MPI_(I)Session_dyn_query_psetop} is used by MPI processes to query pending MPI PSet operations involving a given process set.
This is useful in cases where MPI processes require information about PSet operations related to particular PSets and the PSet operation has been specified by another MPI process or through external mechanisms.
The procedure is collective across all processes in a given MPI process set to enable consistency of queried information across all involved processes.
Listing \ref{lst:mpi_psetop} provides the signature and description for a blocking and a non-blocking version of this procedure.

\begin{lstlisting}[frame=single, basicstyle=\scriptsize, language=C, label={lst:mpi_query_psetop}, escapeinside={<@}{@>},
caption={Procedure for querying information about pending MPI PSet operations}]
int MPI_(I)Session_dyn_query_psetop(
 MPI_Session session,   // session handle
 char *coll_pset,       /* IN: Name of the collective Pset
 			 * - call is collective across the processes in 
 			 *   coll_pset
 	                 * - all processes receive the same query results
 	                 */ 
 char *pset_name,       // IN: Name of the Pset for which the psetop 
 			// should be queried 
 int *op,               // OUT: The MPI PSet operation
 char ***input_psets,   // OUT: Array of input Pset of the operation 
 int *input_psets,      // OUT: Number of input Psets		
 char ***output_psets,  // OUT: Array of output Pset of the operation 
 int *noutput_psets,    // OUT: Number of output Psets
 (MPI_request *request) // OUT: MPI_Request for the non-blocking version
)

<@\textbf{Description:}@> Query <@for@> pending PSet operation invloving the PSet <@\textit{pset\_name}@>.
If no pending PSet operation involves the PSet, <@op@> will be set to 
MPI_PSETOP_NULL. The call is collective over the processes in 
<@\textit{coll\_pset\_name}@>. "mpi://SELF" may be used <@for@> non-collective lookups.

\end{lstlisting}

\textbf{Completing pending MPI PSet operations}:
The procedure \lstinline{MPI_Session_dyn_complete_psetop} is used by MPI processes to acknowledge receipt of information about a pending PSet operation involving the specified process set.
Upon return from this procedure, the PSet operation is considered completed, i.e. it is not pending anymore.
A subsequent call to \lstinline{MPI_(I)Session_dyn_query_psetop} with the same PSet argument will therefore either return \lstinline{MPI_PSETOP_NULL} or a new pending PSet operation involving the specified PSet.
Listing \ref{lst:mpi_psetop} provides the signature and description of this procedure.

\begin{lstlisting}[frame=single, basicstyle=\scriptsize, language=C, label={lst:mpi_finalize_psetop}, escapeinside={<@}{@>}, caption={Procedure for completing an MPI PSet operation.}]
int MPI_Session_dyn_complete_psetop(
  MPI_Session session, // IN: session handle
  char *pset_name      // IN: Name of the Pset for which the psetop
  		       // should be finalized (one of the in/output sets)
)
RETURN: 
   - MPI_SUCCESS <@if@> operation was successful

<@\textbf{Description:}@> Indicates completion of the PSet Operation. This will make the 
operation  unavailable <@for@> <@\textit{MPI\_Session\_dyn\_query\_psetop}@>.

\end{lstlisting}

\textbf{Storing/Retrieving Data associated with MPI PSets}:
The procedures \lstinline{MPI_Session_publish_data} and \lstinline{MPI_Session_lookup_data} can be used by MPI processes to publish and lookup key-value pairs in the data store associated with the provided PSet name.
Listing \ref{lst:mpi_publish} and \ref{lst:mpi_lookup} provide the signatures and descriptions for a blocking and a non-blocking version of these procedures.
\begin{lstlisting}[frame=single, basicstyle=\scriptsize, language=C, label={lst:mpi_publish}, escapeinside={<@}{@>}, caption={Procedure for publishing key-value pairs in the data store associated with a PSet.}]
int MPI_Session_publish_data(_nb)(
 MPI Session session    // IN: Session handle
 const char* pset_name  // IN: Name of PSet in whose Dictionary the data 
 			// should be published
 MPI_Info info          // IN: Info object containg key-value pairs to be 
 			// published
 (MPI_Request *request) // OUT: MPI Request for the non-blocking version
)

<@\textbf{Description:}@> Publishes a key-value pair in the data store associated with 
the given PSet name. The PSet name has to be considered valid by MPI.

\end{lstlisting}

\begin{lstlisting}[frame=single, basicstyle=\scriptsize, language=C, label={lst:mpi_lookup}, escapeinside={<@}{@>}, caption={Procedure for retrieving key-value pairs from the data store associated with a PSet.}]
int MPI_Session_lookup_data(_nb)(
 MPI_Session session,   // IN: Session handle
 char *coll_pset_name,  // IN: Name of a PSet specifying procs of the 
 			//     collective - all processes receive the 
 			//     same value(s)
 	                  
 char *pset_name,       // IN: Name of PSet in whose Dictionary the data 
 			// should be looked up
 char **keys,           // IN: Keys for which values should be looked up
 int nkeys,             // IN: The number of keys
 int wait,              // IN: Flag indicating to wait for values to be 
 			//     published
 MPI_Info *info_used    // OUT: MPI_Info containing requested key-values
 (MPI_Request *request) // OUT: MPI Request for the non-blocking version
)

<@\textbf{Description:}@> Looks up a key-value pair in the data store associated with 
the given PSet name. The PSet name has to be considered valid by MPI. 
The call is collective over the processes in coll_pset_name, i.e., all 
processes in coll_pset_name have to call this function. 
All processes are guaranteed to receive the same values. 
"mpi://SELF" may be used <@for@> non-collective lookups.

\end{lstlisting}

\subsection{PMIx}
PMIx is a standardized interface specification for services in distributed computing systems and is widely used e.g. in MPI implementations to realize the interaction with the MPI runtime.
Therefore, PMIx should provide an interface for dynamic resource services. In the following subsections, we provide insights into the realization of our design principles in PMIx and propose exemplary API extensions.
In the following, we rely on the specifications of the PMIx Standard 5.0 \cite{pmix50}.

\textbf{Principle 1:}
PMIx is based on PMIx Processes.
Thus, according to principle 1 \textit{PMIx processes are used for resource allocation}.
PMIx processes are associated with resources such as CPU cores through mapping/pinning or other means.

\textbf{Principle 2:}
The PMIx Standard 5.0 defines a PMIx PSet as a ``user-provided or host environment assigned label associated with a given set of application processes. The host environment can define and delete process sets.'' This definition is identical to our definition of a process set except for the ordering condition.
Thus, according to principle 2, \textit{the granularity of PMIx process changes should be PMIx process sets.}

\textbf{Principle 3:}

Following principle 3 \textit{PMIx process changes would be described as set operations and relations between PMIx PSets}.

\textbf{Principle 4:}
Since PMIx is a standardized API for accessing services in distributed systems, as a separation of concerns the contents of the COL should not be part of the PMIx specification.
However, PMIx needs to support the transmission of COL objects.

\textbf{Principle 5:}
As principle 5 mandates the association between PSet operations and COL objects, PMIx needs to provide functionality for \textit{associating PMIx PSet operations with COL objects}.
Thus, PMIx needs to be able to handle COL objects passed by the user and associate them with a given PMIx PSet operation.

\textbf{Principle 6:}
As mandated by principle 6 process sets need an associated data storage to allow the association of user data with process sets.
Thus, \textit{PMIx needs to provide functionalities to store and retrieve global data and associate this data with the label of a given PMIx process set}.

\subsubsection{Interface Proposal}
Based on the observations from the previous subsections, we describe one possible way to extend the PMIx interface to support the design principles.
The following interface extensions should not be seen as a proposal but as a source of insight into possible ways of realizing the design principles in PMIx.

\textbf{Specifying PMIx PSet operations}:
PMIx 5.0 provides the \lstinline{PMIx_Allocation_request} API to request a resource allocation action from the host environment.
The current specification does not involve the notion of PSet operations.
Thus, an additional allocation directive (Lst.~\ref{lst:pmix_alloc_dir}) and additional allocation attributes (Lst.~\ref{lst:pmix_alloc_attr}) would be required to support PSet operations via the \lstinline{PMIx_Allocation_request} API.
The host environment would be responsible for spawning processes and defining PSets according to the PSet operation in its implementation of the \lstinline{pmix_server_alloc_fn_t} server function.

\begin{lstlisting}[frame=single, basicstyle=\scriptsize, label={lst:pmix_alloc_dir},caption={Additional allocation directive for PSet Operations}]
PMIX_ALLOC_PSETOP_XXX (pmix_alloc_directive_t):
    A PSet operation is specified, which should be considered in the 
    optimization decision of the host environment. 

\end{lstlisting}
\begin{lstlisting}[frame=single, basicstyle=\scriptsize, label={lst:pmix_alloc_attr},caption={Additional allocation attributes for specifying PSet operations}]
PMIX_ALLOC_PSETOP_INPUT "pmix.alloc.psetop.in" (char *):
    Comma-seperated list of input PSet names for the PSet operation
PMIX_ALLOC_PSETOP_OUTPUT "pmix.alloc.psetop.out" (char *):
    Comma-seperated list of output PSet names for the PSet operation
PMIX_ALLOC_PSETOP_LCLOI "pmix.alloc.psetop.lcloi" (pmix_byte_object_t):
    LCLOI object associated with the allocation request

\end{lstlisting}

\textbf{Querying pending PMIx PSet operations}:
PMIx provides the \lstinline{PMIx_Query_info} API to query information from the host environment.
Thus, the \lstinline{PMIx_Query_info} API could be extended with additional query attributes (Lst.~\ref{lst:pmix_query_attr}) information related to PMIx PSet operations.

\begin{lstlisting}[frame=single, basicstyle=\scriptsize, label={lst:pmix_query_attr}, caption={Proposed additional query attributes}]
PMIX_QUERY_PSETOP_OP "pmix.qry.psetop.op" (pmix_alloc_directive_t):
    Operation of the PSet operation. REQUIRED QUALIFIERS: PMIX_PSET_NAME
PMIX_QUERY_PSETOP_INPUT "pmix.qry.psetop.in" (char *):
    Comma-seperated list of input PSet names of the PSet operation
    REQUIRED QUALIFIERS: PMIX_PSET_NAME
PMIX_QUERY_PSETOP_OUTPUT "pmix.qry.psetop.out" (char *):
    Comma-seperated list of output PSet names of the PSet operation
    REQUIRED QUALIFIERS: PMIX_PSET_NAME

\end{lstlisting}

\textbf{Completing pending PMIx PSet operations}:
PMIx 5.0 provides a notification API to send events and to register event handlers.
Thus, this API could be extended with an additional event code (Lst.~\ref{lst:pmix_event_code}), to allow application processes to indicate the completion of a set operation.

\begin{lstlisting}[frame=single, basicstyle=\scriptsize, label={lst:pmix_event_code}, caption={Proposed additional event code}]
PMIX_EVENT_PSETOP_FINALIZED (pmix_status_t)
    The application intends finalization of the pset operation
    REQUIRED ATTRIBUTS: PMIX_PSET_NAME
\end{lstlisting}

\textbf{Storing/Retrieving Data associated with PMIx PSets}:
PMIx 5.0 provides an API to publish and look up data in a global data store.
However, so far the API does not allow the association of data with specific string tags such as PSet names.
Thus, this API could be extended with a new attribute (Lst.\ref{lst:pmix_data_attr}) and slight revision of the data retrieval rules (Lst.\ref{lst:pmix_data_rev}), to allow an association between PSet names and published data.

\begin{lstlisting}[frame=single, basicstyle=\scriptsize, label={lst:pmix_data_attr}, caption={Proposed additional data attribute}]
PMIX_DATA_TAG "pmix.data.tag" (char *)
    A string tag associated with the data. Data Published with a specific 
    tag can only be retrieved if the same tag is specified in the Lookup 
    operation. Duplicate keys are allowed to exist on the same range 
    if they are associated with different tags.

\end{lstlisting}

\begin{lstlisting}[frame=single, basicstyle=\scriptsize,, label={lst:pmix_data_rev}, caption={Proposed revisions of the PMIx data retrieval rules}]
Revision: PMIx Standard, Version 5.0 (May 2023), p. 120
    Publishing duplicate keys is permitted provided they are published 
    to different ranges <rev> or to the same range with different 
    tags. </rev> Duplicate keys being published on the same data <rev> 
    with the same tag </rev> shall return the PMIX_ERR_DUPLICATE_KEY 
    error.

Revision: PMIx Standard, Version 5.0 (May 2023), p. 131
    2. If the key <rev>(and tag if existant) </rev> of the stored 
    information does not match the specified key <rev> (and tag 
    if specified) </rev>, then the search will continue.
\end{lstlisting}

%% file: chapters/04_discussion.tex
\section{Discussion and Early Experiences}
\label{sec:discussion}

\subsection{Prototype Implementation}
We implemented a prototype mostly based on the interface extensions outlined in the previous section.
Our prototype consists of 4 parts: an extended MPI Sessions interface implemented in Open-MPI, an extended PMIx interface implemented in OpenPMIx, an extended PMIx runtime environment implemented in PRRTE, and a custom resource manager implementation implemented as a Python-based PMIx Tool.
This prototype demonstrates the feasibility of implementing the DPP approach in a commonly used HPC software stack.

All of the mentioned software is publicly available \cite{dynprocs_repo}.

\subsection{Programming effort}
We used the extended MPI Session interface to implement support for dynamic resources in different applications.
These efforts provided two insights:

1. The extended MPI Session interface is very flexible, allowing it to address the peculiarities of applications, and

2. Due to its generic and flexible design, the DPP interface requires a comparably high programming effort to achieve the desired dynamic behavior in applications.

From these observations, we conclude that the DPP interface should be seen as a low-level interface serving as a basis on which more specialized dynamic resource libraries can be built on top.
For instance, we extended P4est \cite{burstedde11} - a commonly used adaptive mesh refinement library - to leverage DPP as the underlying mechanism for dynamic resources, while hiding its complexities from users through a more specialized interface.
Similar library adaptions and specialized interfaces could be provided for different application use cases to ease the programming effort while relying on a generic, standardized mechanism to interact with the RMS.

\subsection{Optimization information}
The resource optimization in the DPP design is based on the association of COL objects associated with the PSet operations specified by applications.
The COL should enable resource managers to make informed decisions based on precise knowledge of the resource requirements and expected performance of applications.
Thus, the COL needs to provide complex information generically and efficiently.
To this end, careful design and standardization of the COL contents is required, which we see as a research area on its own.
The current specification of the interfaces should therefore be seen as a specification of the way applications and resource managers interact, with the content of the COL being an orthogonal concern.

\subsection{Required changes in Resource Management Software}
The shift from a static to a dynamic resource model significantly impacts the requirements for resource management software.
We see three major areas where RMS needs to be adapted:

\textbf{Processing of runtime information:}
The introduction of dynamicity requires RMS to process runtime information in real time.
This includes instance system and application monitoring data as well as optimization information provided by applications such as PSet operations and COL objects.
This implies a significant rise in the amount and frequency of data processing required for optimizing dynamic resource management.

\textbf{Optimization algorithms:}
Current (static) scheduling policies mostly use variations of EasyBackfilling heuristics. These heuristics are based on the job arrival time, the number of requested resources, and the requested time as specified in the batch script.
Dynamic scheduling policies need to be based on more complex optimization algorithms which also take into account scalability and energy models as well as data redistribution overheads to decide on dynamic reconfigurations.
We therefore expect significantly increased computational requirements for dynamic optimization algorithms, possibly on specialized hardware.

\textbf{Decentralization}
The aforementioned increases in data processing and computation required for optimizing dynamic resource management as well as the increasing scale of HPC systems necessitate a shift from the traditional, centralized architecture to a distributed, hierarchical architecture of RMS. This would allow for aggregation and filtering in the data processing as well as model order reduction and parallel evaluation of dynamic optimization algorithms.
Flux \cite{ahn14} is a recent example of RMS with a hierarchical design, addressing the challenges of future large-scale HPC systems.

%% file: chapters/05_related_work.tex
\section{Related Work: Dynamic Resources in Cloud Computing and HPC}
\label{sec:related_work}

In this section, we first present an overview of dynamic resources beyond the conventional HPC execution model, such as cloud computing and workflow execution. We then discuss common shortcomings of dynamic resource approaches for supercomputers and traditional workloads. Finally, we provide an overview of past approaches for dynamic resources in HPC and highlight how these approaches suffer from the discussed shortcomings. The principles we introduced in this work aim to overcome these limitations in highly scalable, high-performance programming models.

\subsection{Differences Between Cloud and HPC and Key Drivers of Dynamicity}

Dynamic resources, often referred to as ``elasticity'' or ``autoscaling'' in the context of cloud computing, serve as a fundamental pillar. Cloud computing diverges from traditional HPC in various ways, including on-demand resource allocation, virtualization, containerization, and workload types. The extensive use of virtualization and containerization techniques in the cloud environment facilitates dynamicity and the seamless migration of cloud applications. This evolution is primarily driven by the need for flexibility in typical cloud workloads, encompassing microservices \cite{joao21}, edge computing, and complex workflows.

As a result of these factors, the literature surrounding dynamicity and elasticity in cloud computing is notably more extensive compared to the HPC domain. Thus, we direct readers to reference \cite{dhuraibi18}, which offers a comprehensive literature overview of the state-of-the-art and research challenges of elasticity in the cloud.

\subsubsection{Taxonomy of elasticity in the cloud}
In this section, we provide a summary of the taxonomy and challenges related to elasticity in the cloud, as outlined in this publication, and draw parallels to dynamicity in HPC.

\cite{dhuraibi18} categorizes elasticity based on criteria such as configuration, scope, purpose, mode or policy, method or action, architecture, and provider:

\textbf{Configuration:} Describes the initial resource reservation process from a cloud provider. Common configurations in the cloud include on-demand, in-advance, best-effort, and action-based reservations. HPC systems predominantly offer best-effort reservations, with exceptions in the realm of urgent computing \cite{yoshimoto12}.

\textbf{Scope:} Indicates the level at which elasticity actions are applied. Infrastructure-level elasticity in the cloud involves scaling virtualization instances, such as VMs or containers, without requiring changes to the application's source code. Embedded elasticity, on the other hand, integrates the elasticity controller directly into the application source code or utilized framework. While infrastructure-level elasticity is prevalent in the cloud due to virtualization techniques and client-server applications, infrastructure-level elasticity in HPC is mostly restricted to I/O malleability and energy/power control. In HPC, achieving changes in allocation size usually necessitates embedded elasticity to enable application adaptation to resource fluctuations.

\textbf{Purpose:} In cloud environments, elasticity serves diverse purposes, balancing cloud provider benefits with Quality of Service (QoS) for users. In HPC, QoS traditionally is a secondary concern after optimizing throughput, resource usage, and power efficiency \cite{flich_16}, however, a trade-off between individual application requirements and overall system optimization is crucial in dynamic resource management strategies.

\textbf{Mode or Policy:} Mode or policy describes the mechanisms that trigger reconfiguration actions. Automatic rescaling is common in the cloud, differentiating between reactive and proactive approaches. Reactive approaches rely on threshold-based triggers for reconfiguration, while proactive approaches anticipate future system or application state changes to initiate proactive rescaling. Techniques like Reinforcement Learning and Control Theory can be applied to both. HPC would benefit from both reactive and proactive approaches, but due to its tightly coupled applications, reconfiguring compute resources often requires direct interaction with applications. This necessitates mechanisms that provide insights into reconfiguration support and its impact on applications.

\textbf{Method:} Methods for reconfiguring cloud applications typically involve horizontal scaling (changing the number of instances, e.g., virtual machines or containers) or vertical scaling (adjusting the resources allocated to a single instance). In contrast, dynamic compute resources in HPC typically focus on altering the number of processes and their associated resources. It therefore invloves aspects from both, horizontal scaling and vertical scaling requiring embedded elasticity within the application source code for tasks like establishing communication with new processes.

\textbf{Architecture:} Cloud elasticity architectures can be centralized or decentralized, each with its advantages. Decentralized approaches distribute the workload across multiple components, increasing scalability but also complexity. HPC resource managers traditionally follow a centralized approach (e.g., SLURM, PBS), while newer designs like Flux explore hierarchical and decentralized approaches. The growth in HPC system scale and resource optimization complexity drives the shift toward decentralized architectures.

\textbf{Provider:} Elasticity solutions in the cloud can target a single cloud provider or multiple providers, spanning elasticity within one provider or across several. In contrast, HPC workload distribution across multiple HPC systems is less common, with dynamic resource approaches typically focused on a single HPC system. However, within this context, approaches can differ based on whether they target a specific resource management software or multiple resource managers.

\subsubsection{Challenges and Open Issues}

Reference \cite{dhuraibi18} outlines several challenges and open issues related to elasticity in the cloud:

\textbf{Interoperability and Unified Platforms:} The need for generic abstractions to ensure interoperability and unified platforms for elastic applications.

\textbf{Granularity of Resources and Billing:} The granularity of resource changes and billing can be too coarse, necessitating more flexible granularity.

\textbf{Advanced Optimization:} Challenges in defining thresholds, managing trade-offs between interests, and addressing prediction errors, as well as hybrid solutions.

\subsection{Workflows in Cloud and HPC}

Scientific workflows have gained immense popularity in recent years, describing the process of achieving scientific outcomes, often represented as directed graphs of computational tasks \cite{ludaescher09}. While these tasks frequently involve typical HPC workloads, scientific workflows are often executed in the cloud due to the elastic cloud execution model. In traditional HPC systems, scientific workflows are typically executed using either job chaining or pilot jobs \cite{rodrigo17}. Job chaining represents workflows as traditional batch jobs with dependencies, optimizing resource utilization but often resulting in longer turnaround times due to waiting times in the scheduling queue. Pilot jobs, on the other hand, involve submitting a single pilot job with sufficient resources and using a workflow manager to execute tasks within the allocated resources. While this approach reduces turnaround time, it may lead to resource wastage. Dynamic resource management approaches are required to address these challenges in HPC.

\subsection{Dynamic resource management in HPC}
\label{sec:related_work_hpc}
In this section, we discuss typical shortcomings of dynamic resource approaches in HPC and give an overview of past approaches in the last two decades.

\subsubsection{Typical shortcomings of dynamic resource approaches in HPC}
\textbf{Programming model or RMS specific approaches:}
Many approaches target only specific programming models or RMS and therefore fail to establish a programming model agnostic standard interface for dynamic resources in HPC.
A programming model agnostic interface is required to facilitate interoperability with the resource management software on the system.

\textbf{Granularity problem:}
A common limitation of many approaches is a too coarse-grained granularity of resource changes.
Many approaches require global synchronization across all processes of a job for a reconfiguration and only target job-level optimization.
This is often due to relying on manipulation of a global communication context, such as the \lstinline{MPI_COMM_WORLD} communicator in MPI.
Besides obvious performance concerns, this also significantly harms the flexibility of these approaches for jobs consisting of different, loosely coupled parts such as task graph-based applications or coupled simulations.
Therefore, a finer granularity than job-level granularity is required for a generic approach for dynamic resources.

\textbf{MPI 2.0 Dynamic Process Management limitations:}
Since MPI 2.0 the MPI Standard provides functionality for Dynamic Process Management (DPM).
A central pillar of MPI DPM are the \lstinline{MPI_Comm_spawn}, \lstinline{MPI_Comm_connect}, \lstinline{MPI_Comm_accept} and \lstinline{MPI_Comm_disconnect} functions. \lstinline{MPI_Comm_Spawn} is collective over an input communicator and provides functionalities to spawn new MPI processes at runtime.
The resulting inter-communicator connects the input communicator with a new, disjoint \lstinline{MPI_COMM_WORLD} communicator which contains the new MPI processes.
Similarly, \lstinline{MPI_Comm_connect} and \lstinline{MPI_Comm_accept} allow the creation of an inter-communicator from two disjoint communicators (possibly derived from disjoint \lstinline{MPI_COMM_WORLD} communicators. 
The MPI 2.0 Dynamic Process Management has several limitations.

One problem is the blocking behavior of \lstinline{MPI_Comm_spawn}. This can lead to significant overheads as processes cannot continue their execution while new processes are launched.

Another problem is the connectedness of MPI processes created by a single invocation of\lstinline{MPI_Comm_spawn} or \lstinline{MPI_Comm_connect}/\lstinline{MPI_Comm_accept} respectively.
While processes from different \lstinline{MPI_COMM_WORLD} communicators can be disconnected with \lstinline{MPI_Comm_disconnect}, this is not possible for processes that share the same \lstinline{MPI_COMM_WORLD}. Therefore, the granularity in which MPI processes can be removed is restricted to the number of processes in the \lstinline{MPI_COMM_WORLD} communicators.
Approaches that try to overcome this issue, e.g. by sequentially spawning single processes, lead to high overheads for process spawning.

Finally, the specification of MPI DPM 2.0 does not describe the interaction with resource management software to change the current allocation.
Most implementations therefore default to spawning new processes inside of the current allocation, potentially oversubscribing resources.

Due to these limitations, approaches based on \lstinline{MPI_Comm_spawn} fall short of the performance and flexibility expected of a dynamic resource approach for HPC.

\textbf{Only either System-driven (malleability) or application-driven (evolving):}
Many approaches provide an interface for either system-driven changes of resources (malleability) or application-driven changes of resources (evolving).
Both approaches cannot reach optimal dynamic resource management due to a lack of local or global optimization information respectively.
Purely System-driven changes such as lack of information about application specifics such as reconfiguration points, data redistribution overheads, different application phases, etc.
Although monitoring data can provide some insights into application behavior, optimization potential is lost due to a lack of information about the dynamic behavior of applications.
On the other hand, a purely application-driven approach, where applications request particular resources from the system lacks information about the global state of the system.
Therefore the system will not be able to optimize the global resource assignments based on such application-driven requests.
To globally optimize resource assignments in the system a cooperative approach between the resource manager and applications is required.

\textbf{Limited applicability to real-world applications:}
So far, many approaches have only been demonstrated to work with particular application types (such as iterative solvers) or strongly simplified applications (benchmarks).
These applications are not representative of the large variety of HPC applications encountered on real systems and hence cannot be deployed successfully on these systems.

\subsubsection{Past approaches for dynamic resources in HPC}
Several approaches for dynamic resources have been developed over the past two decades.
Some of the earliest approaches originate from research on fault tolerance, in particular checkpoint/restart (C/R) mechanisms.
In \cite{vadhiyar03} and \cite{lemarinier16} the authors leverage the Stop Restart Software (SRS) and the Scalable Checkpoint/Restart (SCR) \cite{moody10} libraries respectively for storing/restoring checkpoints and restarting applications with different numbers of processes.
These approaches have the advantage of relatively low code-intrusiveness and applicability to a large range of applications.
However, these approaches exhibit the granularity problem, as it requires restarting the whole application.
Moreover, there is a significant overhead involved with the relaunch of the application processes and the restoration of application data from stored checkpoints.
Finally, these approaches do not provide means for applications to provide local optimization information to the system, hence it is a purely system-driven approach.

In \cite{maghraoui07} the authors provide an API for writing malleable, iterative MPI applications leveraging the Process Checkpoint and Migration (PCM) \cite{maghraoui06} library.
The API provides means to migrate, split, or merge MPI processes relying on checkpointing for data redistribution.
This approach avoids a complete application restart by reconfiguring the global \lstinline{MPI_COMM_WORLD} communicator.
However, reconfiguration requires global synchronization of all processes.
Moreover, the approach is specific to iterative MPI applications.

The Charm++ \cite{kale93} parallel programming model has been the basis of several projects related to dynamic resources in HPC.
Charm++ is based on over-decomposition and migratable objects and therefore allows for adaptivity and load balancing.
Adaptive MPI (AMPI) \cite{huang04} is a mostly standard compliant MPI implementation on top of Charm++ and supports migration of MPI ranks and automatic load balancing.
In \cite{kale02}, the authors presented a malleable job system consisting of an adaptive Charm++ runtime and an adaptive job scheduler, which supports changing the processor set of AMPI and Charm++-based applications.
In another publication \cite{prabhakaran15} the Maui/Torque Batch System has been adapted to support scheduling malleable and evolving jobs and combined with the adaptive Charm++ runtime to change the resources of Charm++-based jobs accordingly.
While these projects are very promising, they are specialized on the Charm++ programming model and runtime and a particular batch system implementation.
Moreover, the provided interface only has very limited support for fully cooperative resource management.
Here, we see our approach as complementary, providing a generic mechanism for dynamic resource management which could be leveraged by Charm++/AMPI to improve its interoperability with other RMS and to support optimization of complex applications.

ReSHAPE \cite{sudarsan07} is a framework for dynamic resizing and scheduling of homogeneous applications. It consists of a custom scheduling and monitoring module and a programming model for resizing applications.
The user API allows for transmission of a small set of application-local optimization information which is gathered to create application profiles as a basis for resize decisions.
Therefore, ReSHAPE provides some support for collaborative optimization, targeting in particular the type of homogeneous, iterative MPI applications.
As the process reconfiguration is based on providing a custom, malleable \lstinline{MPI_COMM_WORLD} communicator where the process creation is based on MPI 2.0 DPM, it also suffers from the aforementioned shortcomings inherent in these approaches.

DMR API \cite{iserte20} and DMRlib \cite{iserte21} provide APIs for writing iterative, malleable applications targeting the OmpSs and MPI programming models respectively, while running under the Slurm resource manager.
Users provide functions for data redistribution and periodically check for required reconfiguration actions. Upon reconfiguration, DMR automatically grows or shrinks the global DMR communicator using MPI 2.0 Dynamic Process Management and redistributes the data using the user-provided data redistribution functions.
DMR allows for easy programming of malleable, iterative OmpSs / MPI applications.
However, as it is based on MPI 2.0 DPM it exhibits the mentioned limitations of this approach.
Moreover, it only targets particular programming models, RMS, and application types and therefore does not provide a generic interface for dynamic resources.

Elastic MPI \cite{compres16} provides extensions of the MPI programming model and adaptions of the SLURM infrastructure to allow system-driven reconfiguration of the \lstinline{MPI_COMM_WORLD} communicator to adapt to changing resources.
However, Elastic MPI requires global synchronization of all processes to reconfigure the global \lstinline{MPI_COMM_WORLD} communicator, hence limiting the granularity.
Moreover, the approach is specific to the MPI programming model and only supports purely system-driven reconfiguration, lacking local optimization information.

FlexMPI \cite{gonzalo13} provides infrastructure for dynamic load-balancing based on profiling of MPI applications.
In \cite{martin15} FlexMPI was extended with functionalities for dynamic reconfiguration of the number of processes based on performance models deduced from profiling data.
The reconfiguration mechanism of FlexMPI is based on a custom, malleable \lstinline{MPI_COMM_WORLD} communicator, where MPI 2.0 DPM is used for creating new processes and therefore also inherits the granularity problem and MPI 2.0 DPM shortcomings.

Overall, past approaches have focused on providing dynamic resource mechanisms using various techniques. However, all of these approaches share some common shortcomings due to focusing on introducing these mechanisms only in particular applications, programming models, and RMS without addressing the need for more abstract, generic interfaces for dynamic resources in HPC programming models.

%% file: chapters/06_summary.tex
\section{Summary}
\label{sec:summary}
Dynamic Resource Management is considered to be an important research area to improve the efficiency of resource utilization in modern HPC systems.
However, while elasticity is already a widely adopted technique in cloud computing, similar techniques have not yet been established on most production HPC systems.
A major reason is that enabling dynamic resources in HPC typically requires deeper embedding of dynamic resource mechanisms in application source code, whereas elasticity in the cloud can often be achieved on the virtualization layer, such as scaling the number of container instances.
Many publications presented approaches to overcome this challenge over the last two decades, however, none of these approaches has been widely adopted so far.
We account for this due to a lack of a generic, standardized interface between applications and the RMS, which requires a careful co-design between the application and system perspective.

In this work we assumed resources to be accessible by processes, separating Dynamic Process Management (DPM) and Dynamic Resource Management (DRM).
Applications express their capabilities for dynamism as abstract process management operations on sets of processes.
The RMS then decides on concrete instantiation of such operations and their mapping on the system resources.
These decisions need to be based on both, local application-specific optimization information and global, system-wide optimization information to holistically optimize for a collective optimization objective.

To this end, we proposed six general design principles which in our opinion are mandatory when designing such an interface.
We then discussed a possible concrete realization of these design principles using the Message Passing Interface (MPI) and the Process Management Interface - Exascale (PMIx).
We further shared our experiences with an actual prototype implementation.
We conclude that our design principles could help overcome typical shortcomings of past approaches, paving the way toward a standardized, holistic interface for dynamic resource management for high-performance parallel programming models.

%% file: chapters/99_appendix_history.tex
\section{Sources of inspiration}
\label{sec:appendix}

The ideas presented in this work originate from discussions and experiences across several years and research groups.
Thus in this section, we like to share our source of inspiration.

A major influence was the work carried out in the Invasive Computing project \cite{teich11}, which explored a new paradigm for the design and resource-aware programming of future parallel computing systems.
The project addressed dynamic resources across multiple layers of the embedded system hard- and software stack and produced valuable insights into the challenges involved.
This is also where some ideas of the LOC originated from.

Simultaneously, the MPI community started to discuss new ways to improve the modularity and runtime awareness of MPI, resulting in the creation of the MPI Sessions Working Group.
Many ideas presented in this work stem from discussions in this working group.

Finally, we also learned from our previous work \cite{fecht22,huber22} in which we introduced early versions of a dynamic resource interface.
The design principles presented in this work are the result of the discussions and feedback we gathered around these previous versions.